\documentclass[prb,twocolumn,amsmath,amssymb,aps,superscriptaddress]{revtex4-2}

\usepackage{graphicx}
\usepackage{epstopdf}
\usepackage{dcolumn}
\usepackage{bm}
\usepackage{color}

\begin{document}
\title{Characteristic times for gap relaxation and heat escape in nanothin NbTi superconducting filaments: thickness dependence and effect of substrate}
\author{K. Harrabi}
\affiliation{Physics Department, King Fahd University of Petroleum and Minerals, 31261 Dhahran, Saudi Arabia}
\affiliation{Interdisciplinary Research Center (RC) for Intelligent Secure Systems, KFUPM, Dhahran 31261, Saudi Arabia }
\author{A. Mekki}
\affiliation{Physics Department, King Fahd University of Petroleum and Minerals, 31261 Dhahran, Saudi Arabia}
\author{M. V. Milo\v{s}evi\'c}
\affiliation{Department of Physics, University of Antwerp, Groenenborgerlaan 171, B-2020 Antwerp, Belgium}

\begin{abstract}
We measured the temporal voltage response of NbTi superconducting filaments with varied nanoscale thicknesses to step current pulses that induce non-equilibrium superconducting states governed by a hot-spot mechanism. Such detected voltage emerges after a delay time t$_{d}$, which is intimately connected to the gap relaxation and heat escape times. By employing time-dependent Ginzburg-Landau theory to link the delay time to the applied current, we determined that the gap relaxation time depends linearly on film thickness, aligning with the acoustic mismatch theory for phonon transmission at the superconductor-substrate interface. We thereby find a gap relaxation time of 104 ps per nm of thickness for NbTi films on polished sapphire. We further show that interfacial interaction with the substrate significantly impacts the gap relaxation time, with observed values of 9 ns on SiO$_{x}$, 6.8 ns on fused silica, and 5.2 ns on sapphire for a 50 nm thick NbTi strip at $T = 5.75$~K. These insights are valuable for optimizing superconducting sensing technologies, particularly the single-photon detectors that operate in the transient regime of nanothin superconducting bridges and filaments. 	
\end{abstract}
\date{\today}

\pacs{}
\maketitle

\section{Introduction}

Both non-equilibrium superconductivity and thermal transport in superconducting materials are of pertinent importance over the past decade for their direct relevance to established and emerging nano-electronic devices. For example, single-photon detectors based on superconducting meanders readily found their way to practical applications in optical communications and quantum technology \cite{Zhong}. 

It is well known that superconducting filaments can carry electrical current without dissipation as long as the current is kept below the critical value I$_{c}$, beyond which the non-equilibrium state is reached and the energy dissipates via formation of phase-slip centers and hot spots \cite{Tinkham1}. The formation of such resistive states results in occurrence of well defined resistance steps in the measured current-voltage ($I$-$V$) characteristics \cite{Webb,Meyer}. For example, corresponding voltage jumps were reported in the $I$-$V$ characteristics of ultra-thin superconducting NbN nanowire grown on a sapphire substrate~\cite{Delacour}, where at low temperatures a crossover between the thermal and the quantum behaviour in the phase-slip regimes was thoroughly studied. Quantum phase slip centers were also investigated in a superconducting NbN wire of width reduced down to several superconducting coherence lengths ($\xi$) \cite{Constantino}, and distinction has been made between the coherent and incoherent quantum phase slip events, the latter being fostered by thermally-activated fluctuations.

On the theory end, the time-dependent Ginzburg-Landau (TDGL) theory is known to be the most effective formalism to capture the resistive state of current-carrying superconductors \cite{Tinkham1}. For example, this methodology was successfully used to describe the current-induced resistive state in 1D \cite{Vodolazov2005,Elmurodov2008}, 2D \cite{Berdiyorov2009,Berdiyorov2009_2,Silhanek2010,Berdiyorov2012,Berdiyorov2014}, and 3D \cite{Berdiyorov2008,Liu2011,Berdiyorov2009_3,Berdiyorov2013,Wang2017} superconductors. This approach can also account for other sample-specific features such as pinning centers \cite{Berdiyorov2011,Berdiyorov2012_2,Berdiyorov2012_3,Latimer2013,Berdiyorov2015}, disorder and spatial inhomogeneities \cite{Lombardo}, external \cite{Berdiyorov2013,Berdiyorov2009_4} and internal \cite{Berdiyorov2009_5,Berdiyorov2012_4,Berdiyorov2017} magnetic field, which all affect the dynamics of the superconducting state under current bias, and exhibited a very good agreement with most experiments to date \cite{Elmurodov2008,Silhanek2010,Berdiyorov2012_3,Latimer2013,Berdiyorov2014,Berdiyorov2015,Wang2017,Lombardo}. 

On the technology end, creation of a localized non-equilibrium state (i.e., hot spot) under driving current is the working mechanism of the Superconducting Nanowire Single-Photon Detectors (SNSPDs) \cite{Vodolazov1,Berdiyorov2012_5}. In Ref.~\onlinecite{Ferrari2017} such hot spot was created in NbN waveguide-integrated superconducting nanowire under current bias and irradiation of photons, and hot-spot relaxation time was deduced. The heat dissipated in the localized hot spot is evacuated towards the substrate by phonons. The thermal boundary conductance between the superconducting film and the substrate quantifies the phonon rate emission and heat evacuation via the substrate. The quality of such thermal interfacing is a dominant feature that affects the performance and sensitivity of devices such as SNSPDs, as well as their jitter time, switching current, and efficiency~\cite{Allmaras}. The thermal boundary conductance between superconducting nanowires and different substrates were also measured using the hotspot current and the mismatch model~\cite{Dane}. These studies showed the importance of the substrate to the out-of-equilibrium dynamics of the superconducting state in current-carrying superconducting systems. 

In this work, we analyze the dependence of the dynamic properties of NbTi superconducting bridges under pulsed current on the thickness of the superconducting filament and the quality of its interface with the substrate. We established that the delay time t$_{d}$ before transitioning to the resistive state decreases with increasing current pulse amplitude. Using time-dependent Ginzburg-Landau theory, we then determine that the gap relaxation time depends linearly on film thickness, for a given substrate, consistent with the acoustic mismatch theory. We then proceed to investigate the non-equilibrium states in NbTi filaments on different substrates, finding that substrate material and the interface quality significantly impact gap relaxation and heat escape times of the superconducting nanostructure. These findings highlight the critical importance of film thickness and substrate choice for optimizing superconducting electronics. The ability to threby control the relaxation time is essential for the development of high-performance devices in applications such as single-photon detectors, quantum computing, and advanced superconducting circuits, where precise manipulation of superconducting states is of paramount importance.  

\section{Experimental setup and detection of the expected dissipative states}
\begin{figure}[t]
\includegraphics[width=\linewidth]{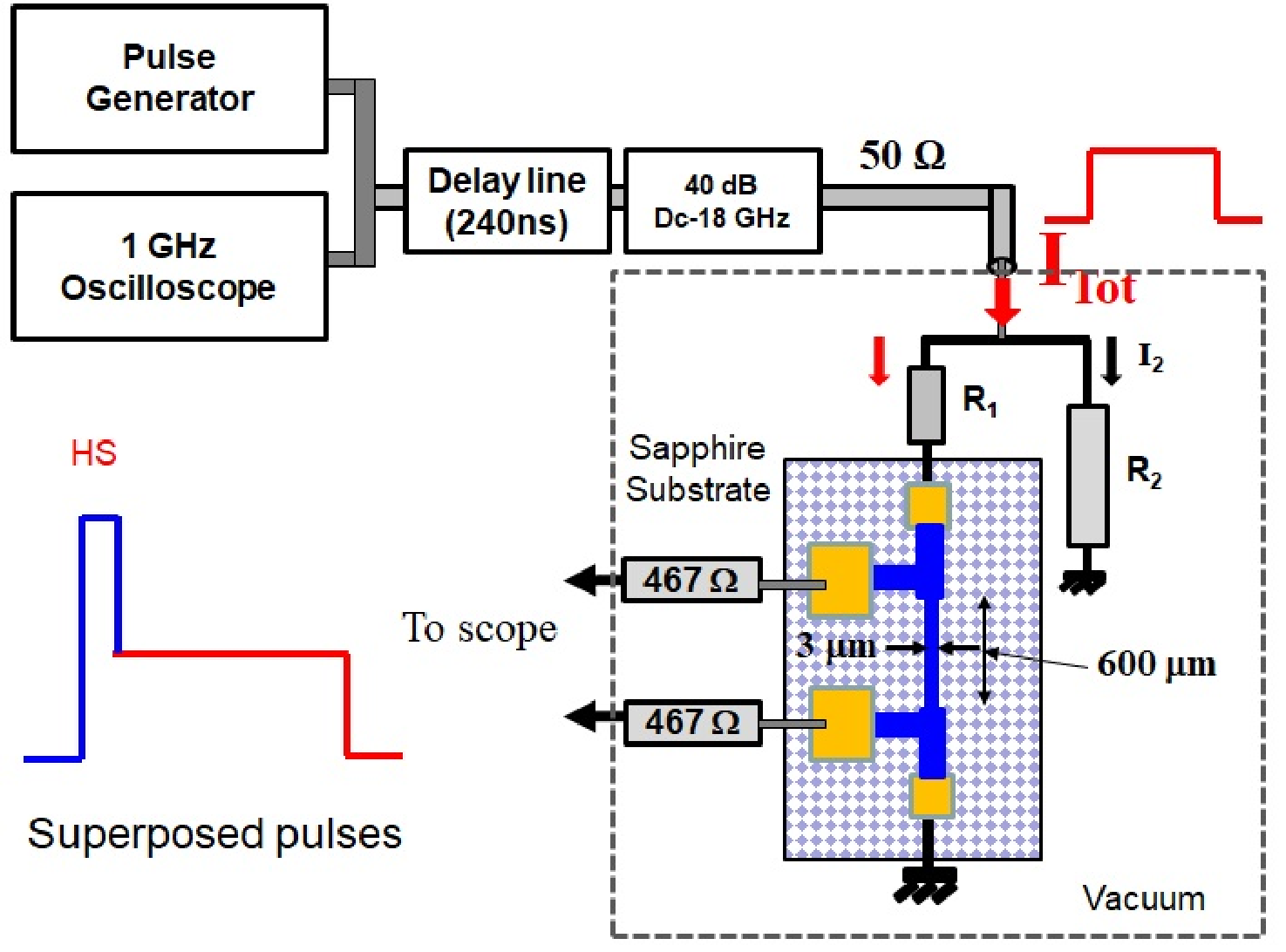}
\caption{Schematics of the experiment: the lateral probes are connected to the oscilloscope through 467 $\Omega$ resistors to avoid any current deviation to the outside circuit. The sample is mounted in parallel to a resistor R$_{2}$ to ensure that constant current is fed through the sample. Two current pulses were superposed and sent to the filament: the shortest pulse is used to create a hot spot (HS), and the second one with the plateau serves to maintain it.}
\label{fig1}
\end{figure}

In this experiment, Niobium-Titanium (NbTi) wires with a width of 3.0 $/mu$m and varied thicknesses (30, 50, and 80 nm) were deposited on a sapphire substrate via sputtering in an Ar-N plasma under vacuum conditions (STAR-Cryoelectronics, NM, USA). The superconducting wires and electrical connections were created using photolithography and ion milling, resulting in a sample layout featuring a 600 $\mu$m long wire for current pulse application, as shown in Fig. \ref{fig1}. Voltage measurements were taken across the wire using a double-sided probe set. Additionally, 50 nm thick NbTi wires with a width of 5 $\mu$m were fabricated and deposited on sapphire, fused silica, and silicon oxide substrates. Current pulses were applied to the wire, which was grounded at the other end, with voltage measurements taken across the wire. The measurements were conducted under vacuum, using current pulses of variable amplitude, 450 ns duration, and 10 kHz repetition rate, applied via a 240 ns air-delay line. Voltage was monitored with a fast oscilloscope, and the samples exhibited superconducting transition temperatures around 8.0 K.

In a current transport measurement, the zero resistance survives up to a critical current I$_{c}$, beyond which dissipation arises through local onset of phase-slip lines \cite{Harrabi1, Lyatti} and/or dissociation of the vortex-antivortex pairs. Such phase slips and vortices exhibit high velocity \cite{Embon}, dependent on applied magnetic field. The generated quasi-particles diffuse over an inelastic length, cause heating and dissipation, and a clear voltage signal emerges. 
Within a phase-slip line (PSL), the collapse and restoration of the order parameter occurs at the Josephson frequency, and the phase of the order parameter changes by 2$\pi$ across the PSL. In a current-biased transport measurement, the normal current within a PSL is the source of Joule heating, leading to hot spots and nucleation of local normal zones above a certain thermal threshold current I$_{h}$, dependent upon the cooling conditions. Depending on sample specifics, the two currents I$_{c}$(T) and I$_{h}$(T), have different dependencies, and the crossing temperature T* between these two currents defines the nucleation temperature of a PSL, usually confined to a restricted temperature range close to T$_{c}$. However, below T* a hot spot regime is found. Some thin films showed a completely different behaviour where the PSL regime extends over the entire range of temperature below T$_{c}$ \cite{Harrabi2}. 

Therefore, a superconducting bridge biased with a current exceeding the switching value can present two different thermal behaviours. In these two states the dissipated energy leads to attaining a temperature smaller or larger than T$_{c}$. Such dissipation was discussed in superconducting bridges near T$_{c}$, where the theoretical study and numerical calculations showed that in both thermal and quasi-equilibrium limits the delay time t$_{d}$ is independent of temperature, which contradicts with the non-thermal model that showed a strong temperature dependence \cite{Berdiyorov2009_2,Vodolazov2005}. Therefore in this work, we revisit the issue both experimentally and theoretically, and use the observed behaviour of the delay time to extract the characteristic dissipative times for thin superconducting filaments on different substrates - as relevant for applications in superconducting electronics and sensing.

\begin{figure}[t]
\includegraphics[width=\linewidth]{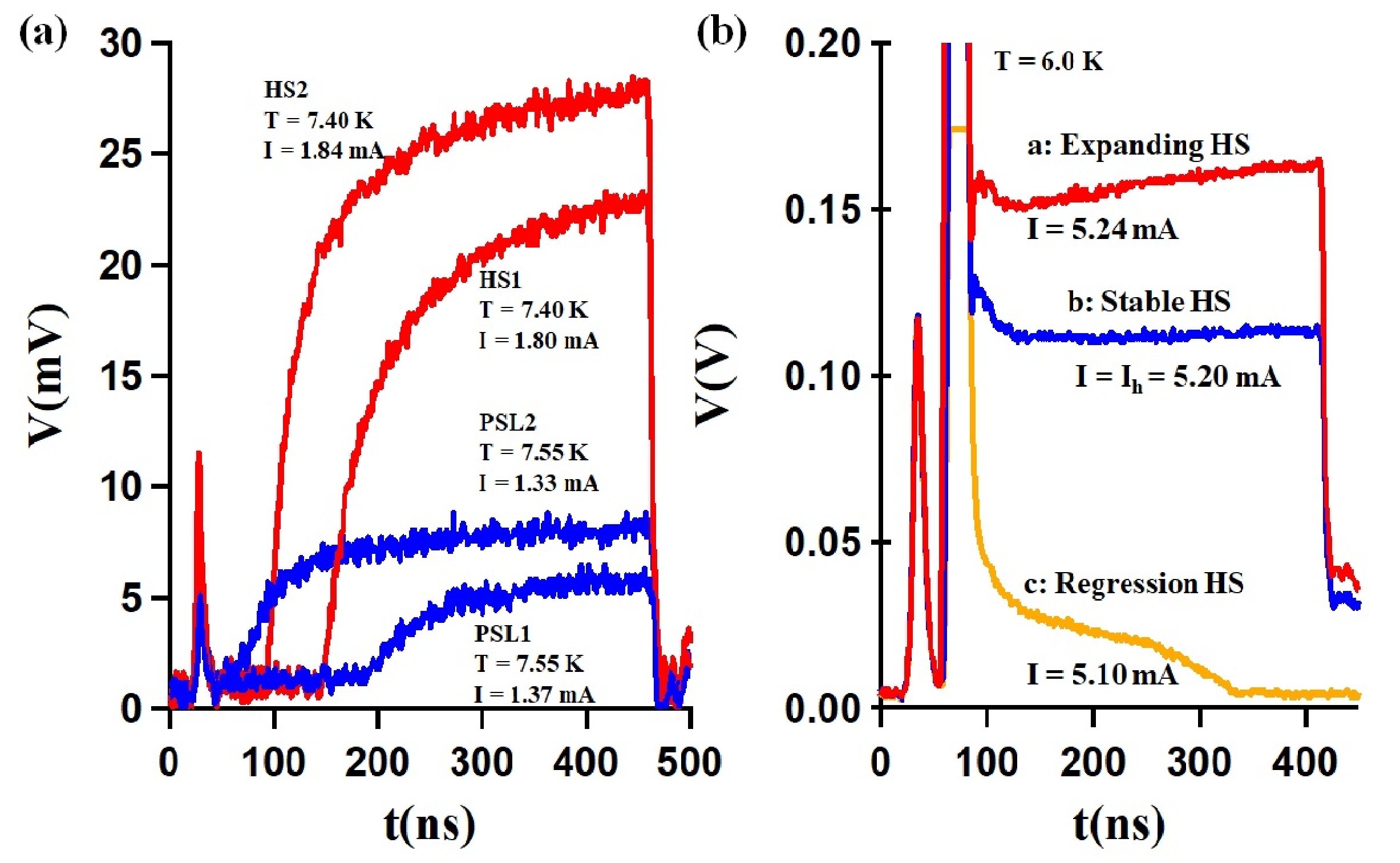}
\caption{ (a) Voltage responses of the two distinct dissipative regimes of the 80 nm thick NbTi bridges. At T = 7.4 K $<$ $T^{*}$ the hot spot (HS) is created (HS1 and HS2 correspond to applied currents I$_{HS2}$ $>$ I$_{HS1}$). At T = 7.55~K $>$ T$^{*}$ the phase-slip line (PSL) is nucleated (PSL1 and PSL2 correspond to applied currents I$_{PSL2}$ $>$ I$_{PSL1}$). (b) The response to two step current pulses at $T = 6$~K, one causing formation of the HS, while the second supported HS in the expansion (I $>$ I$_{h}$), stable (I $=$ I$_{h}$), or regression mode (I $<$ I$_{h}$). }
\label{fig2}
\end{figure} 

\section{Creation of a hot spot using an electrical current pulse}

In a superconducting nanowire, with width comparable to $\xi$, biased with current slightly below its critical current, any thermal excitation (such as one caused by photon absorption) will cause the transition to the normal state. A localized excitation such as impact of a photon will locally induce a normal zone and oblige the current to flow around it, which causes current density to exceed the critical value J$_{c}$ around the initial normal zone and so cascade the entire width of the sample into a hot spot. The measured peak voltage is then used a marker for the detection of the induced resistive state. A similar formation of a hot spot can be obtained by using an electrical current pulse. In this case, a voltage appears after a certain delay time t$_{d}$, after superconductivity locally collapses and a non-equilibrium resistive zone emerges. This zone is brought to temperature larger than the critical temperature, where quasi-particles are generated. The traces in Fig.~\ref{fig2}(a) show the formation of such a hot spot (HS) and the nucleation of a phase-slip line (PSL) in our experiment, respectively away from T$_{c}$ and in the vicinity of T$_{c}$, for the 80 nm thick NbTi bridge. The formation of a HS is accompanied by a monotonically increasing voltage after t$_{d}$ while the PSL exhibits step-like behaviour with a voltage attaining a plateau at times beyond t$_{d}$. To discriminate between the two dissipative modes, and to measure the hot-spot current, two superposed pulses were used. The first one with a duration of 50 ns and a current amplitude set slightly above I$_{c}$, aimed to generate a hot spot. The second pulse with variable amplitude was used to stabilize the HS and avoid its expansion. Fig.~\ref{fig2}(b) shows the voltage response to two superposed pulses, the HS created by the first pulse followed by a linear voltage increase for the rest of the remaining time for a current larger than I$_{h}$.
By lowering the current amplitude during the second pulse, one can produce a steady hot spot and thus define a threshold hot spot current. The trace $b$ of Fig.~\ref{fig2}(b) presented a stable voltage, that marks the value of the threshold current for a stable hot spot, yielding a constant voltage response. However, if one reduces further the amplitude of the second current pulse, one will reach the hot-spot regression regime (trace $c$ of Fig.~\ref{fig2}(b)). Conversely, larger amplitude of the second pulse may support the expansion of the hot spot (trace $a$ of Fig.~\ref{fig2}(b)), but the range of current amplitudes for such expansion is limited.

\begin{figure}[t]
\includegraphics[width=\linewidth]{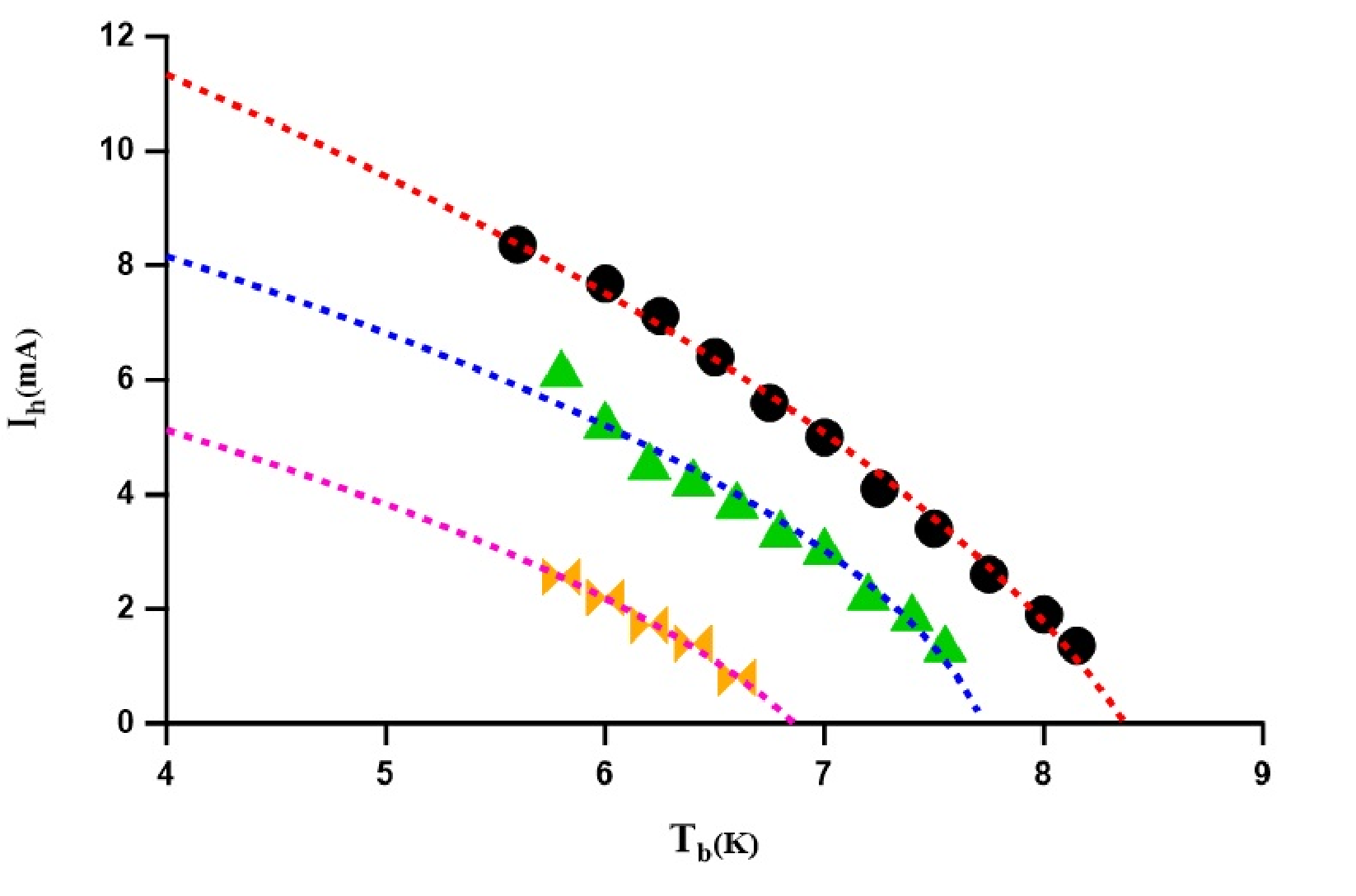}
\caption{The threshold hotspot current, measured using two-step current pulses, versus the substrate temperature, for different thickness of the NbTi bridge (30, 50 and 80 nm). The experimental data close to T$_{c}$ are well fitted with the (1-T/T$_{c}$)$^{0.5}$ dependence, in accordance with theoretical predictions in Ref.~\onlinecite{Skocpol}. }
\label{fig3}
\end{figure}
By systematically repeating the above procedure at different temperatures and monitoring the dissipative processes as discussed in Fig.~\ref{fig2}, we measured and mapped out the temperature dependence of the threshold hot-spot current for three NbTi filaments of different thickness, as presented in Fig.~\ref{fig3}. The obtained values are in excellent functional agreement with the model proposed in Ref.~\onlinecite{Skocpol}, with the energy dissipated as R$_{N}$I$_{h}^{2}\propto$(T-T$_{c}$), i.e. I$_{h}$(T)$\propto$(T-T$_{c}$)$^{1/2}$.

\section{Analysis of delay time and gap relaxation time}

The delay time t$_{d}$ is regarded as the time needed to locally destroy the superconductivity upon external current action. The time-dependent Ginzburg-Landau theory stipulates that t$_{d}$ depends on the normalized order parameter $f$, the ratio between the applied and the critical current, and a prefactor $\tau_{\Delta}$, as
\begin{equation}
t_{d}(I/I_{c})=\tau_{\Delta}\int_{0}^{1} \frac{2f^{4}df}{\frac{4}{27}(\frac{I}{I_{c}})^{2}-f^{4}+f^{6}}.
\label{eq:Eq1}
\end{equation}
Previous studies showed that delay time is related to the energy gap relaxation time \cite{Kardakova}. When Cooper pairs are dissociated, quasi-particles are excited to high energy levels; they relax to the bottom of the energy band where the recombination process takes place and the heat is evacuated to the substrate. $\tau_{\Delta}$ is identified as the gap relaxation time of the filament. 
\begin{figure}[t]
\includegraphics[width=\linewidth]{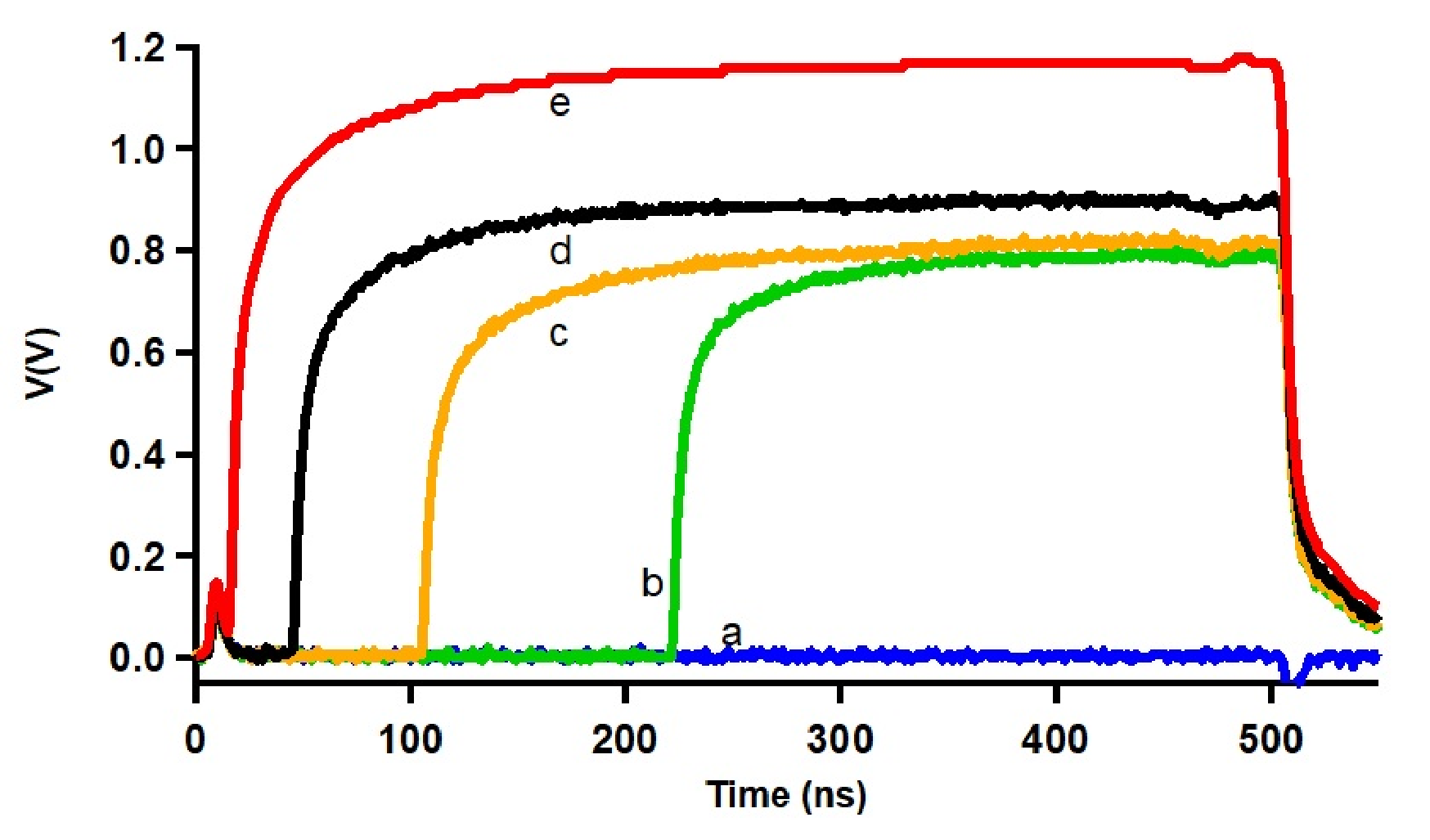}
\caption{Temporal evolution of the voltage induced by a single step current pulse in a 80 nm thick NbTi bridge at T = 5.75~K. Voltage onsets after a certain delay time t$_{d}$, with t$_{d}$ reduced with increasing applied current $I/I_c =$ 1, 1.02, 1.07, 1.12, and 1.25 (traces labeled $a$-$e$ respectively). }
\label{fig4}
\end{figure}
Fig.~\ref{fig4} reveals the reduction of the delay time t$_{d}$ as the applied current is increased, which is intuitively expected. This film cooling time is correlated with the phonon escape time. The evolution of energy dissipation after the creation of the non-equilibrium hot spot also has $\tau_{\Delta}$ as a characteristic time. It implicates the contribution of phonons and electrons, in contrast to the evacuation of the heat into the substrate where only phonons are involved for the period of time $\tau_{esc}$. We thus have $\frac{\int c_{e} dT + \int c_{p} dT }{\tau_{\Delta}} = \frac{ \int c_{p} dT }{\tau_{esc}}$, where $c_{e}$ and $c_{p}$ are the specific heat of the electrons and phonons respectively. The heat escape time can be deduced from the previous equation as $\tau_{esc} = \tau_{\Delta} c_{p} /(c_{p}+c_{e}$). The specific heat values $c_{e}$ and $c_{p}$ in our case can be considered close to the Nb values~\cite{Ladan}. Taking the same ($c_{e}$+$c_{p}$) /$c_{p}\simeq3$, as in Nb, we obtain $\tau_{esc}\simeq\tau_{\Delta}/3$.

\subsection{Thickness-dependence of gap relaxation time}

The interface between the superconducting film and the substrate affects the type of the non-equilibrium dissipative state created in the superconducting film. Among other effects, such boundary interface strongly affects the phonon behaviour. Hence, beside its ever-present effects in a transport measurement, an interface also stirs the thermal conductivity. On that note, while earlier study of nanowires illustrated the drop of the thermal conductivity at a rough interface, that is not observed for a smooth interface~\cite{R.Chen2008}. The theoretical model developed by Rothwarf and Taylor~\cite{Rothwarf}, showed that phonons with energies larger than twice the superconducting gap had an influence on the generated quasi-particles. In such a case, multiple subsequent pair destruction and recombination processes cause the increase of the quasiparticles population. Consequently, the lifetime of the quasi-particle becomes longer with increasing the acoustic mismatch between the film and the substrate, which slows the phonon escape process from the film towards the substrate. The recombination time of the quasi-particles depends on the thickness of the film ($d$) and the pair-breaking mean free path ($l_{ph}$). If $d \geq l_{ph}$, the effective time is susceptible to the film thickness, conversely for $d\leq l_{ph}$ it is independent of the thickness. In case of single-photon detection it is therefore of practical importance to improve the acoustic mismatch between the film and the substrate to reduce the heating effect~\cite{Kaplan}.

The heat generated by a current exceeding the critical value increases the population of the phonons in the localized zone and scales with the thickness. Different combination processes contribute to the energy relaxation rate, notably the electron-phonon scattering, the recombination phenomena, and the electron impurity scattering. The first allows the change of the quasi-particle energy via the interaction with phonons. However, in the second process the condensation of quasi-particle into Cooper pairs is associated with phonon emission. The acoustic mismatch between the two lattices plays a role in reducing the transmission factor $\eta$, but such variation of the transmission coefficient may be omitted in the present analysis since all films of different thickness were evaporated using the same technique and on the same substrate. However, the dominant mechanism is the one with large phonon population, where the phonon-phonon interactions escalate and phonons impede each other, causing the phonon transport to became diffusive rather than ballistic. That slows down the evacuation of the heat from the film to the substrate and causes elongation of the heat escape time from the film to the substrate. 

\begin{figure}[t]
\includegraphics[width=\linewidth]{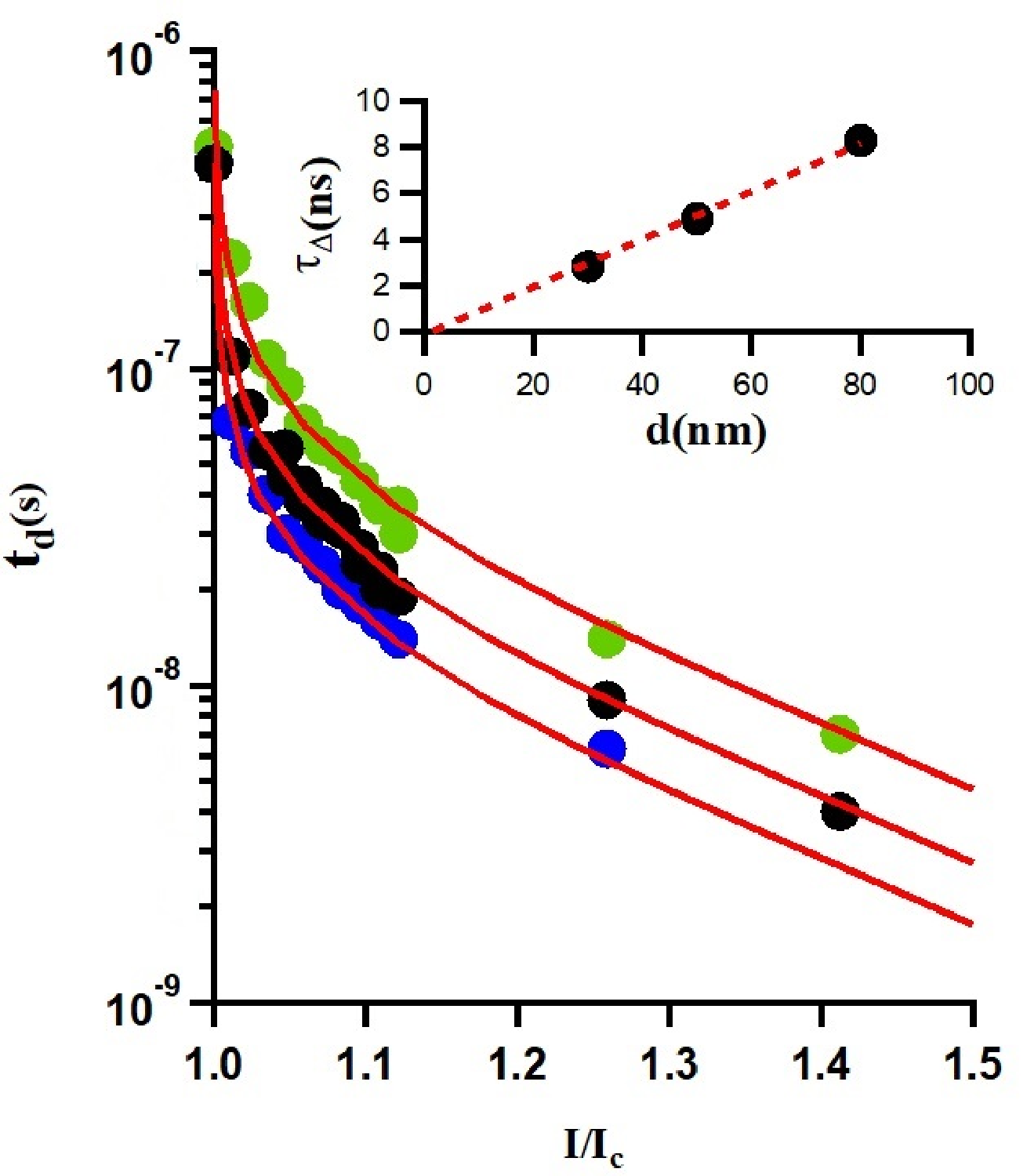}
\caption{Delay times (in log scale) as a function of the applied reduced current I/I$_{c}$, at T $= 3.6~K$ (extracted from Fig.~\ref{fig4}). The red curve is the TDGL functional behaviour given by Eq.~\eqref{eq:Eq1}, yielding determination of $\tau_{\Delta}$= (2.8 $\pm$ 0.2) ns, (4.9 $\pm$ 0.2) ns, and (8.3 $\pm$ 0.3) ns, for three considered thicknesses of the sample. The inset shows the corresponding escape time $\tau_{esc}\approx \tau_\Delta/3$ (see discussion in the text).}
\label{fig5}
\end{figure}

Within the scope of the heat transfer between the superconducting material and the substrate one can conveniently rely on the acoustic mismatch model~\cite{Kaplan}. For simplicity, we assume that the superconducting films and their substrates could be approximated as isotropic solids. As mentioned earlier, the heat escape time depends on the phonon mean free path $l_{ph}$, the film thickness, and the transmission coefficient $\eta$. The mean escape time can then be approximated in case of $d\geq l_{ph}$ by: 
\begin{equation}
\tau_{esc} = 4d/(u\eta) \propto d,
\label{eq:Eq2}
\end{equation}
where $u$ is the velocity of sound in the film. The $\tau_{esc}$ depends solely on thickness in the present investigation. The estimate of the heat escape time from the TDGL approach is in good agreement with the acoustic mismatch model, as the experimental data shown in (the inset of) Fig.~\ref{fig5} exhibit a linear variation of the gap relaxation time with the film thickness $\tau_\Delta/4d\approx 26$~ps/nm ($\tau_{esc}/4d\approx 8.6$~ps/nm). Based on the classical isotropic acoustic mismatch model, the values of $\tau_{esc}/4d$ were estimated from different measurement techniques and reported for some materials and substrates in the past, but this data is rather scarce. For example, Kardakova \textit{et al.} reported electron-phonon relaxation time $\tau_{e-ph}= 2$~ns for 80 nm thick TiN film, and the heat escape time of the order of few ns~\cite{Kardakova}. A full analysis of the electron-phonon and heat escape times exceeds the scope of the present paper, being strongly dependent on the properties of the films, such as its transition temperature, resistivity, and critical current density, all of which are dependent on the film quality.

\subsection{Effect of substrate on characteristic time scales}

\begin{figure}[t]
\includegraphics[width=\linewidth]{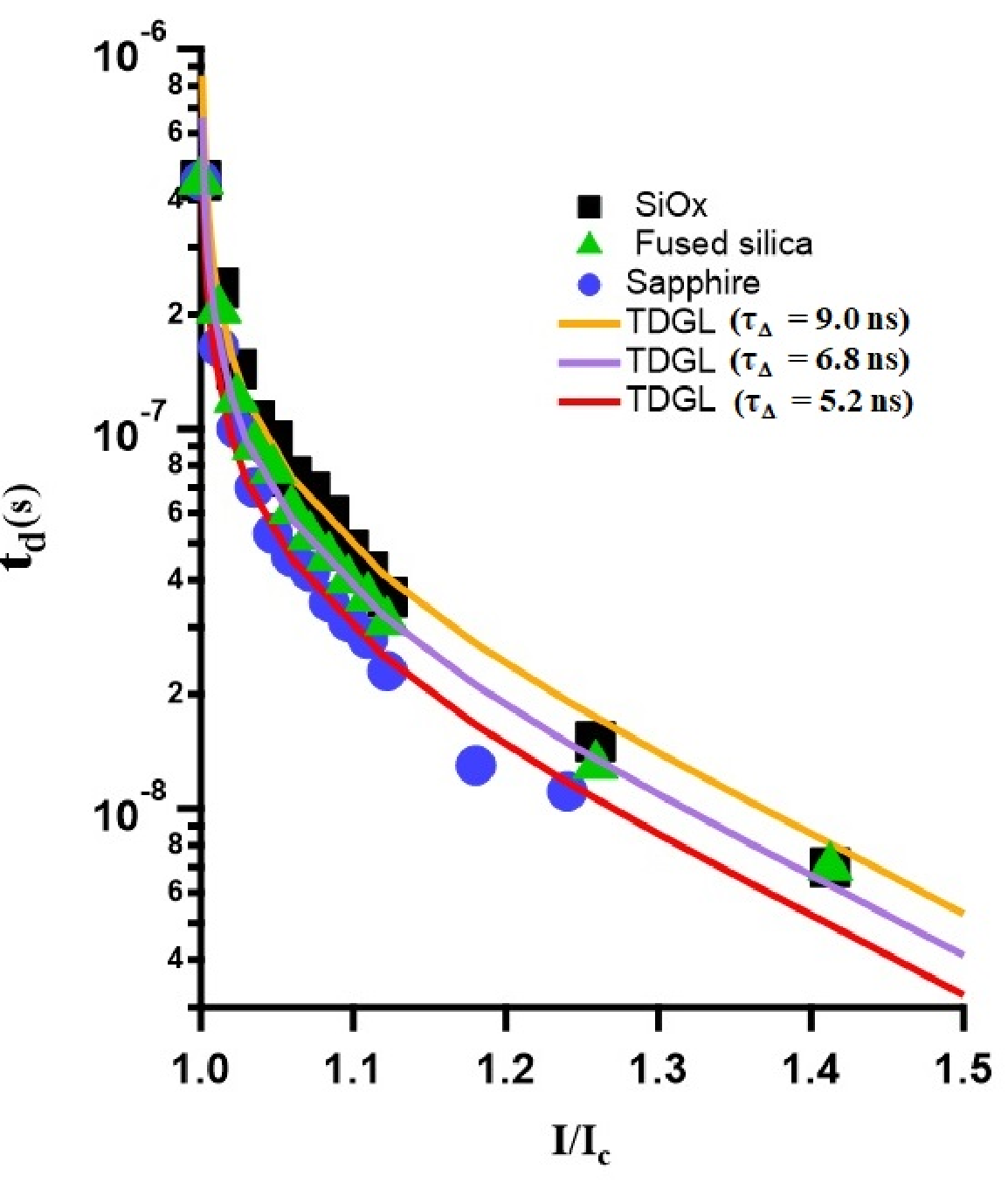}
\caption{Delay time versus applied current characteristics of a 50 nm thick and 5 $\mu$m wide NbTi filament deposited on SiO$_{x}$ (black squares), fused silica (green triangles) and sapphire (blue circles), at temperature T = 5.75 K. The solid lines are the fitting curves using Eq.~\eqref{eq:Eq1}, with a prefactor $\tau_{\Delta}$ = (5.2 $\pm$ 0.2) ns, (6.8 $\pm$ 0.2) ns, and (9.0 $\pm$ 0.2) ns for sapphire, fused silica, and SiO$_{x}$ substrates, respectively.}
\label{fig6}
\end{figure}

Having measured the thickness dependence of the characteristic dissipative times in a NbTi superconducting filament on a given substrate, we extend our investigation to filaments of same thickness (in this case 50~nm) deposited on three different substrates. The delay times extracted from NbTi filaments on either sapphire, fused silica, and SiO$_x$ substrates are plotted with their corresponding current values in Fig.~\ref{fig6}, and fitted with the TDGL model using equation \eqref{eq:Eq1}, so that the characteristic time preceding the integral, the gap relaxation time $\tau_\Delta$, can be extracted. The thereby obtained gap relaxation times are $\tau_{\Delta}$ = 5.2 ns, 6.8 ns, and 9 ns for the NbTi filaments on SiO$_{x}$, fused silica, and sapphire, respectively.

The temperature reached at the core of the dissipative zone exceeds the substrate temperature~\cite{Harrabi2}. As a result, heat is generated in a normal zone positioned between two superconducting sinks and is transferred to adjacent areas via phonons and quasi-particles, which leads to the expansion of the HS region. At the same time, heat is also conducted to the substrate, mainly through phonon interactions. As previously discussed, the non-equilibrium system can be divided into two subsystems, quasi-particles and phonons. The Rothwarf-Taylor model was used to describe the time evolution and the dynamics of these two subsystems. The energy dissipation in the non equilibrium zone occurs with characteristic time $\tau_{\Delta}$. This mechanism of dissipation evolves both electron-electron and electron-phonon interactions. However, this generated heat is evacuated towards the substrate and is assured by the phonon-phonon interactions. This phenomenon is thus in tight dependence with the quality of the interface between the superconducting filament and the substrate. 

In our case, the gap relaxation time for the film on sapphire was found to be shorter compared to films on fused silica and SiO$_x$ substrates. The sapphire substrate, having a single crystal structure, contrasts with the amorphous structures of the fused silica and SiO$_x$ substrates. The morphology of the substrate at the interface significantly influences the phonons responsible for dissipating heat into the substrate sink. In the case of sapphire, diffusive phonons primarily facilitate heat transfer, whereas for fused silica and SiO$_x$, ballistic phonons dominate.

Exploring the heat transfer between the superconducting material and the substrate involves using the acoustic mismatch model~\cite{Rothwarf}. Ballistic phonons, characterized by coherent and uninterrupted motion, and diffusive phonons, which exhibit more random and scattered movement, both affect the efficiency of heat dissipation at the interface. Understanding these phonon characteristics is crucial for optimizing heat dissipation strategies in superconducting systems with various (substrate) interfaces, especially in modern circuit designs. When examining heat transfer at the interface between a superconducting material and a substrate, thermal conductance ($G$) is an important factor. Thermal conductance is given by $G =A\frac{k_{s}\cdot k_{sc}}{k_{s}+k_{sc}}$, where $k_{s}$ is the thermal conductivity of the substrate, $k_{sc}$ is the thermal conductivity of the superconducting material, and A is the cross-sectional area of the interface. This equation accounts for the combined thermal conductivities at the interface, providing insights into heat transfer efficiency.

For ballistic phonons, another critical parameter is their mean free path ($l_{ph}$), which reflects phonon behaviour with longer mean free paths, indicating less frequent collisions and more coherent motion. This coherent motion impacts heat transport across the interface, affecting overall heat transfer efficiency. At temperatures well below the superconducting transition temperature (T$_c$), the thermal conductivity of the superconducting film is primarily dominated by phonons, making the interface quality and phonon mean free path crucial.

In cryogenic environment, sapphire's higher phonon transmission typically results in more favorable thermal boundary conductance. For superconducting films used in high-performance electronics or quantum computing, a substrate with higher thermal conductivity like sapphire aids in better heat removal.

For layered structures or devices where thermal management is critical, understanding and optimizing interfacial thermal conductance is essential. Techniques like Time-Domain Thermoreflectance (TDTR) can measure and optimize these properties. Sapphire, with better thermal conductivity and longer mean free paths compared to fused silica and SiO$_x$, has a thermal conductance relationship 
G$_{sapphire}$ $<$ G$_{fused silica}$ $<$ G$_{SiOx}$. This aligns with our measured heat escape times, where sapphire shows faster escape times compared to fused silica, and SiO$_x$ exhibits the slowest one. This order of escape times, $\tau_{sapphire} < \tau_{fused silica} <  \tau_{SiOx}$, summarizes the distinct thermal dynamics of these substrates.

Assuming superconducting films and substrates as isotropic solids, and sufficiently thick films, the gap relaxation and heat escape time will depend on thickness as stipulated by Eq.~\ref{eq:Eq2}. Using the same filament thickness in our analysis of the effect of the substrate, the transmission coefficient $\eta$ remains as the only variable. Thereby estimated values of $u\eta$ for NbTi on sapphire, fused silica and SiO$_x$ are 12.82, 9.80 and 7.41 m/s, respectively. This in turn yields the thickness dependence of the characteristic heat escape time as $\tau_{esc}=312$~ps per nm, $408$~ps per nm, and $540$~ps per nm, for NbTi on sapphire, fused silica and SiO$_x$, respectively. This information is of high practical value towards optimized design of superconducting sensing devices based on heat absorption and general optimization of superconducting electronics where excess heat has detrimental effects that need be controlled if not removed.

\section{Conclusion}
To extract the characteristic time scales of gap and heat relaxation in thin superconducting filaments, we conducted pulsed-current transport measurements on 2D NbTi superconducting micro-bridges of different thickness and on different substrates, to investigate their temporal response to overcritical current, and their behaviour in the resistive state. For that goal, we first established the threshold current for a stable resistive state, and mapped out its dependence on temperature. Then, for a given temperature, we detected the temporal delay of the suppression of superconductivity, as a function of applied current. After corroborating the data with the Time-Dependent Ginzburg-Landau (TDGL) model, we determined the gap relaxation and heat escape time, which display a linear correlation with film thickness, aligning with predictions from the acoustic mismatch model. Specifically, thin NbTi filaments on sapphire exhibited a fast gap relaxation time of 104 ps per nm of thickness, suggesting suitability of such nanostructures for advanced superconducting electronics. We extended our investigation into gap relaxation time across various substrates (sapphire, fused silica, and SiO${x}$), to highlight the crucial influence of the interface with the substrate. We report slower heat escape times for SiO${x}$ and fused silica substrates compared to sapphire, attributed to different phonon transport mechanisms and amorphous interface. These results and insights underscore the importance of substrate selection and interface quality in optimizing the performance of superconducting detectors and electronic devices, which is pivotal e.g. for advancing superconductor-based single-photon detectors and bolometers.

\section*{ACKNOWLEDGMENTS}
The authors gratefully acknowledge the support of the King Fahd University of Petroleum and Minerals, Saudi Arabia, under the DF191008 DSR project.

\end{document}